\documentclass[twocolumn]{revtex4}
\usepackage{epsf,graphicx}
\usepackage{amsmath,amssymb,amsfonts}
\begin{document}

\author{D.H.E.~Gross}

\affiliation{{$^1$}Hahn-Meitner-Institut, Bereich Theoretische
Physik, Glienickerstr. 100, 14109 Berlin (Germany)}

\title{No dinosaurs in microcanonical gravitation:\\
No special "microcanonical phase transitions"}

\begin{abstract}It is shown that the recently introduced
singularities of the {\em microcanonical} entropy like
"microcanonical phase transitions", and exotic pattern of the {\em
microcanonical} caloric curve $T(E)$ like multi-valuednes or the
appearance of "dinosaur's necks" are inconsistent with Boltzmann's
fundamental definition of entropy $S=\ln[W(E)]$ for a system at
equilibrium even for extremely large systems as astro-physical
ones.
\end{abstract}
\maketitle

In a recent paper Bouchet and Barr\'{e}\cite{bouchet03} introduced a
special class of "microcanonical" phase transitions. These are defined by
singularities of the microcanonical entropy or the microcanonical
temperature. However, such singularities do not exist in the microcanonical
thermodynamics, i.e. of a system at equilibrium, based on Boltzmann's
principle \cite{gross174} neither for small systems nor for the real large
self-gravitating astrophysical ones.

The entropy as defined by Boltzmann's principle
\begin{equation}S(E)=\ln[W(E)]\end{equation}
as also the temperature
\begin{equation}\beta(E)=T^{-1}(E)=\frac{d S}{dE}\end{equation}
are {\em single valued, smooth, and multiple differentiable} for any finite
$N$. Here $W(E)$ is the ($6N-1$) folded integral which measures the
geometric area of the microcanonical many-fold of points with the energy
$E$ in the $6N$-dim. phase space. $T(E)$ is a single valued differentiable
function at all energies as long as $\beta>0$, i.e. $E$ is finite. This
clearly excludes all microcanonical "hysteretic cyclings", -zones, or
"dinosaur's necks" as introduced by \cite{chavanis02b}.

Chavanis argues in private communication that the "hysteretic cycles"
reflect possible meta-stable configurations which are important in
astro-physics. This might well be so but is outside of any controllable and
fundamental conception of statistics. The microcanonical ensemble ${\cal
E}$ is per definition the ensemble of all possible configuration in the
$6$N-dim phasespace with the same energy. Its geometrical size is
$W=e^{S(E)}$. Of course it is interesting and important to consider only
those configurations ${\cal M}\subseteqq {\cal E}$ which represent e.g.
stars at densities up to the ignition of their hydrogen burning, when
further gravitational collapse is halted for a while. This can be done by
adding an upper limit to the local density as done e.g by ref.
\cite{gross187,gross195,gross200}. ${\cal M}(E)$ is then still a compact
manifold and similarly single valued multiple differentiable in $E$. Again
there are no singularities and no "dinosaurs".

\end{document}